\documentstyle[aps,twocolumn,prl,epsf]{revtex}
\begin{document}

\draft
\input{epsf}

\newcommand{\hz}{\widehat{\mbox{\boldmath{$\zeta$}}}}
\newcommand{\btau}{\mbox{\boldmath{$\tau$}}}
\newcommand{\bxi}{\mbox{\boldmath{$\xi$}}}
\newcommand{\bS}{\mbox{\boldmath{$S$}}}
\newcommand{\bz}{\mbox{\boldmath{$\zeta$}}}
\newcommand{\bJ}{\mbox{\boldmath{$J$}}}
\newcommand{\JO}{\mbox{\boldmath{$J^{0}$}}}
\newcommand{\cD}{{\cal D}}
\newcommand{\cJ}{{\cal J}}
\newcommand{\dashed}{\mbox{-\; -\; -\; -}}
\newcommand{\dotted}{\mbox{${\mathinner{\cdotp\cdotp\cdotp\cdotp\cdotp\cdotp}}$}}
\newcommand{\full}{\mbox{------}}

\title{Cryptographical Properties of Ising Spin Systems}
\author{Yoshiyuki~Kabashima$^{1}$, Tatsuto Murayama$^{1}$ and
David~Saad$^{2}$} \address{$^{1}$ Department of Computational
Intelligence and Systems Science, Tokyo Institute of Technology,
Yokohama 2268502, Japan.  \\ $^{2}$The Neural Computing Research
Group, Aston University, Birmingham B4 7ET, UK.}

\maketitle

\begin{abstract}
The relation between Ising spin systems and public-key cryptography is
investigated using methods of statistical physics. The insight gained
from the analysis is used for devising a matrix-based cryptosystem
whereby the ciphertext comprises products of the original message
bits; these are selected by employing two predetermined
randomly-constructed sparse matrices. The ciphertext is decrypted
using methods of belief-propagation. The analyzed properties of the
suggested cryptosystem show robustness against various attacks and
competitive performance to modern cyptographical methods.

\end{abstract}

Public-key cryptography plays an important role in many aspects of
modern information transmission, for instance, in the areas of
electronic commerce and internet-based communication. It enables the
service provider to distribute a public key which may be used to
encrypt messages in a manner that can only be decrypted by the service
provider. The on-going search for safer and more efficient
cryptosystems produced many useful methods over the years such as RSA
(by Rivest, Shamir and Adleman), elliptic curves, and the McEliece
cryptosystem to name but a few.

In this Letter, we employ methods of statistical physics to study a
specific cryptosystem, somewhat similar to the one presented by
McEliece\cite{McEliece}. These methods enable one to study the typical
performance of the suggested cryptosystem, to assess its robustness
against attacks and to select optimal parameters.

The main motivation for the suggested cryptosystem comes from previous
studies of Gallager-type error-correcting
codes\cite{Gallager,MacKay,Sourlas} and their physical
properties\cite{us_gallager,us_sourlas}. The analysis exposes a
significantly different behaviour for the two-matrix based codes (such
as the MN code\cite{MacKay}) and single-matrix codes\cite{Sourlas},
which may be exploited for constructing an efficient cryptosystem.

In the suggested cryptosystem, a plaintext represented by an
$N$ dimensional Boolean vector $\bxi \in (0,1)^N$ is encrypted to the $M$
dimensional Boolean ciphertext $\bJ$ using a predetermined Boolean matrix $G$,
of dimensionality $M\times N$, and a corrupting $M$
dimensional vector $\bz$, whose elements are 1 with probability 
$p$ and 0 otherwise, in the following manner
\begin{equation}
\label{eq:ciphertext}
\bJ = G \ \bxi \ + \bz \ ,  
\end{equation}
where all operations are (mod 2). The matrix $G$ and the probability
$p$ constitute the public key; the corrupting vector $\bz$ is chosen at
the transmitting end. The matrix $G$, which is at the heart of the
encryption/decryption process is constructed by choosing two
randomly-selected sparse matrices $A$ and $B$ of dimensionality
$M\!\times \!N$ and $M\!\times\! M$ respectively, defining
\[ G \! =\! B^{-1}A \ \  \mbox{(mod 2)} \ . \]
The matrices $A$ and $B$ are generally characterised by $K$ and $L$
non-zero unit elements per row and $C$ and $L$ per column
respectively; all other elements are set to zero. The finite, usually
small, numbers $K$, $C$ and $L$ define a particular cryptosystem; both
matrices are known only to the authorised receiver.
Suitable choices
of probability $p$ will depend on the maximal achievable rate for the
particular cryptosystem as discussed below.

The authorised user may decrypt the received ciphertext $\bJ$ by
taking the (mod 2) product $B \bJ = A\bxi\! + \!  B\bz$.  Solving the
equation
\begin{equation}
\label{eq:decoding}
A\bS + B\btau =A\bxi + B\bz \  \  \mbox{(mod 2)}, \
\end{equation}
is generally computationally hard. However, decryption can be carried
out for particular choices of $K$ and $L$ via the iterative methods of
Belief Propagation (BP)\cite{MacKay}, where pseudo-posterior
probabilities for the decrypted message bits, $P(S_{i}\!=\!1 | \bJ) \
1\!\le\!i\! \le\! N $ (and similarly for $\btau$), are calculated by
solving iteratively a set of coupled equations for the conditional
probabilities of the ciphertext bits given the plaintext and vice
versa.  For details of the method used and the explicit equations see
\cite{MacKay}.

The unauthorised receiver, on the other hand, faces the task of
decrypting the ciphertext $\bJ$ knowing only $G$ and $p$.  The
straightforward attempt to try all possible $\bz$ constructions is
clearly doomed, provided that $p$ is not vanishingly small, giving
rise to only a few corrupted bits; decomposing $G$ to the matrices $A$
and $B$ is known to be a computationally hard problem\cite{NP}, even if the
values of $K,C$ and $L$ are known. Another approach to study the
problem is to exploit the similarity between the task at hand and the
error-correcting model suggested by Sourlas\cite{Sourlas}, which we
will discuss below.

The treatment so far was completely general. We will now make
use of insight gained from our analysis of
Gallager-type\cite{us_gallager} and Sourlas\cite{us_sourlas}
error-correcting codes to suggest a specific cyptosystem construction
and to assess its performance and capabilities.  The method used in
both analyses \cite{us_gallager,us_sourlas} is based on mapping the
problem onto an Ising spin system Hamiltonian, in the manner discovered
by Sourlas\cite{Sourlas}, which enables one to analyse
typical properties of such systems.

To facilitate the mapping we employ binary representations $(\pm1)$ of
the dynamical variables $\bS$ and $\btau$, the vectors $\bJ$, $\bz$
and $\bxi$, and the matrices $A$, $B$ and $G$, rather than the Boolean
$(0,1)$ ones.

The {\em binary} ciphertext $\bJ$ is generated by taking products of
the relevant binary plaintext message bits $J_{\left\langle i_{1},
i_{2} \ldots \right\rangle} \! = \!  \xi_{i_{1}} \xi_{i_{2}} \ldots
\zeta_{\left\langle i_{1}, i_{2} \ldots \right\rangle}$, where the
indices $i_{1},i_{2}\ldots $ correspond to the non-zero elements of
$B^{-1}A$, and $\zeta_{\left\langle i_{1}, i_{2} \ldots
\right\rangle}$ is the corresponding element of the corrupting vector
(the  indices ${\left\langle i_{1}, i_{2} \ldots
\right\rangle}$ corresponds to the specific choice made for each
ciphertext bit).  As we use statistical mechanics techniques, we
consider both plaintext ($N$) and ciphertext ($M$) dimensionalities to
be infinite, keeping the ratio between them $N/M$ finite. Using the
thermodynamic limit is quite natural here as most transmitted
ciphertexts are long and finite size corrections are likely to be
small.

An authorised user may use the matrix $B$ to obtain
Eq.(\ref{eq:decoding}). To explore the system's capabilities one
examines the Gibbs distribution, based on the Hamiltonian
\begin{eqnarray}
\label{eq:Hamiltonian}
{\cal H} &=& \!\!\!\!\! \sum_{<i_1,..,i_K;j_1,..,j_L>}
      \mbox{\hspace*{-5mm}} \cD_{<i_1,..,i_K;j_1,..,j_L>} \ \delta
      \biggl[-1 \ ; \ \cJ_{<i_1,..,i_K;j_1,..,j_L>} \nonumber \\
      &\cdot & S_{i_1}\ldots S_{i_K} \tau_{j_1}\ldots\tau_{j_L}
      \biggr] - \frac{F_s}{\beta} \sum_{i=1}^{N} S_i -
      \frac{F_{\tau}}{\beta} \sum_{j=1}^{M} \tau_j \ .
\end{eqnarray}
The tensor product $\cD_{<i_1,..,i_K;j_1,..,j_L>}
\cJ_{<i_1,..,i_K;j_1,..,j_L>}$, where $\cJ_{<i_1,..,j_L>} \! = \!
\xi_{i_{1}} \xi_{i_{2}}.. \xi_{i_{K}} \zeta_{j_{1}} \zeta_{j_{2}}
.. \zeta_{j_{L}}$, is the binary equivalent of $A\bxi \!  + \!
B\bz$, treating both signal ($\bS$ and index $i$) and the corrupting
noise vector ($\btau$ and index $j$) simultaneously.  Elements of the
sparse connectivity tensor $\cD_{<i_1,..,j_L>}$ take the value 1 if
the corresponding indices of both signal and noise are chosen (i.e.,
if all corresponding elements of the matrices $A$ and $B$ are 1) and 0
otherwise; it has $C$ unit elements per $i$-index and $L$ per
$j$-index, representing the system's degree of connectivity.  The
$\delta$ function provides $1$ if the selected sites' product
$S_{i_1}..S_{i_K} \tau_{j_1}..\tau_{j_L}$ is in disagreement
with the corresponding element $\cJ_{<i_1..j_L>}$, recording an
error, and $0$ otherwise. Notice that this term is not frustrated, and
can therefore vanish at sufficiently low temperatures ($T \!=\!
1/\beta \!\rightarrow\! 0$), imposing the restriction of
Eq.(\ref{eq:decoding}), while the last two terms, scaled with $\beta$,
survive.  The additive fields $F_s$ and $F_{\tau}$ are introduced to
represent our prior knowledge on the signal and noise distributions,
respectively.

The random selection of elements in $\cD$ introduces disorder to the
system which is treated via methods of statistical physics. More
specifically, we calculate the partition function ${\cal Z}
({\cD},\mbox{\boldmath $J$}) = \mbox{Tr}_{\{\bS,\btau\}} \exp [-\beta
{\cal H}]$, which is then averaged over the disorder and the
statistical properties of the plaintext and noise, using the replica
method\cite{us_gallager,Wong_Sherrington}, to obtain the related free
energy ${\cal F} = - \langle \ln {\cal Z} \rangle_{\xi,\zeta,{\cal
D}}$. The overlap between the plaintext and the dynamical vector
$m\!=\!\frac{1}{N}\sum_{i=1}^N \xi_i S_i$ will serve as a measure for
the decryption success.

Studying this free energy for the case of $K\!\!=\!\! L\!\!=\!\! 2$ and in
the context of error-correcting codes\cite{us_gallager}, indicates the
existence of paramagnetic and ferromagnetic solutions depicted in the
inset of Fig.1. For corruption probabilities $p\!>\!p_{s}$ one obtains
either a dominant paramagnetic solution or a mixture of ferromagnetic
($m\!=\!\pm 1$) and paramagnetic ($m\!=\! 0$) solutions as shown in
the inset; thin and thick lines correspond to higher and lower free
energies respectively, dashed lines represent unstable
solutions. Lines between the $m\!=\!\pm 1$ and $m\!=\! 0$ axes
correspond to sub-optimal ferromagnetic solutions.
Reliable decryption may only be obtained for $p\!<\!p_{s}$, which
corresponds to a spinodal point, where a unique ferromagnetic solution
emerges at $m\!=\! 1$ (plus a mirror solution at $m\!=\!  -1$).

The most striking result is the division of the complete space of
$\bS$ and $\btau$ values to two basins of attraction for the
ferromagnetic solutions, for $p<p_{s}$, implying convergence from {\em
any} initialisation of the BP equations.  Critical corruption rate
values for $M/N=2$ were obtained from the analysis and via BP
solutions as shown in Fig.1, in comparison to the rate obtainable from
Shannon's channel capacity\cite{Shannon} (solid line).  The priors
assumed for both the plaintext (unbiased in this case, $F_s=0$) and
the corrupting vector ($F_\tau=(1/2) \ln [(1-p)/p]$) correspond to
Nishimori's condition \cite{Nishimori}, which is equivalent to having
the correct prior within the Bayesian framework\cite{Sourlas_EPL}

The initial conditions for the BP-based decryption were chosen almost
at random, with a very slight bias of ${\cal O}(10^{-12})$ in the
initial magnetisation, corresponding to typical statistical
fluctuation for a system size of $10^{24}$.  Cryptosystems with other
$K$ and $L$ values, e.g., $K,L \ge 3$, seem to offer optimal
performance in terms of the corruption rate they accommodate
theoretically, but suffer from a decreasingly small basin of
attraction as $K$ and $L$ increase. The co-existence of stable
ferromagnetic and paramagnetic solutions implies that the system will
converge to the undesired paramagnetic solution\cite{us_gallager} from
most initial conditions which are typically of close-to-zero
magnetisation. It may still be possible to use successfully specific
matrices with higher $K$ and $L$ values (such as
in\cite{kanter_saad}); however, these cannot be justified
theoretically and there is no clear adventage in using them.

To conclude, for the authorised user, the $K\!\!=\!\! L\!\!=\!\!  2$
cryptosystem offers a guaranteed convergence to the plaintext
solution, in the thermodynamic limit $N\rightarrow \infty$, as long as
the corruption process has a probability below $p_{s}$.  The main
consequence of finite plaintexts would be a decrease in the allowed
corruption rate with little impact on the decoding success.

The task facing the unauthorised user, {i.e.}, 
finding the plaintext from Eq. (\ref{eq:ciphertext})
was investigated via similar methods
by considering the Hamiltonian
\[
{\cal H}\!=\!-\!\!\!\! \sum_{\left\langle i_{1},
..i_{K'} \right\rangle} \!\!\! {\cal G}_{\left\langle i_{1},
..i_{K'} \right\rangle} \ J_{\left\langle i_{1},..i_{K'}
\right\rangle} \ S_{i_{1}}\! .. \! S_{i_{K'}} - \frac{F_s}{\beta} 
\sum_{k=1}^{N} S_{k} \ ,
\]
using Nishimori's temperature $\beta=(1/2) \ln [(1-p)/p]$.  The number
of plaintext bits in each product is denoted $K'$, $\mbox{\boldmath
$S$}$ is the $N$ dimensional binary vector of dynamical variables and
${\cal G}$ is a dense tensor with $C'$ unit elements per index
(setting the rest of the elements to zero) and is the binary
equivalent of the Boolean matrix $G$\cite{us_sourlas}. The latter,
together with the statistical properties of the corrupting vector
$\bz$ constitutes the public key used to determine the 
ciphertext $\mbox{\boldmath $J$}$. The last term on the right is
required in the case of sparse or biased messages and will require
assigning a certain value to the additive field $F_s$.

The matrix $G$ generated in the case of $K\!\!=\!\! L\!\!=\!\!  2$ is dense
and has a certain distribution of unit elements per row.  The fraction
of rows with a low (finite, not of ${\cal O} (N)$) number of unit
elements vanishes as $N$ increases, allowing one to approximate this
scenario by the diluted Random Energy Model\cite{REM} studied
in\cite{us_sourlas} where $K',C'\rightarrow \infty$ while
keeping the ratio $C'/K'$ finite.

To investigate the typical properties of this (frustrated) model, we
calculate again the partition function and the free energy by
averaging over the randomness in choosing the plaintext, the
corrupting vector and the choice of the random matrix $G$ (being
generated by a product of two sparse random matrices). To assess the
likelihood of obtaining spin-glass/ferromagnetic solutions, we
calculated the free energy landscape (per plaintext bit - $f$) as a
function of overlap $m$.  This can be carried out straightforwardly
using the analysis of \cite{us_gallager}, and provides the energy
landscape shown in Fig.2. This has the structure of a golf-course with
a relatively flat area around the one-step replica symmetry breaking
(frozen) spin-glass solution and a very deep but extremely narrow
area, of ${\cal O} (1/N)$, around the ferromagnetic solution. To
validate the use of the random energy model we also added numerical
data ($+$, with error-bars), obtained by BP, which are consistent with
the theoretical results.

This free-energy landscape may be related directly to the marginal
posterior $P(S_{i}\!=\!1 | \bJ) \ 1\!\le\! i\! \le\! N $ and is therefore
indicative of the difficulties in obtaining ferromagnetic solutions
when the starting point for the search is not infinitesimally close to
the original plaintext (which is clearly highly unlikely).  It is
plausible that any local search method, starting at some distance from
the ferromagnetic solution, will fail to produce the original
plaintext.  Similarly, any probabilistic method, such as simulated
annealing, will require an exponentially long time for converging to
the $m\!=\!1$ solution. Numerical studies of similar energy landscapes show
that the time required increases exponentially with the system
size\cite{Parisi}.

Most attacks on this cryptosystems, by an unauthorised user, will face
the same difficulty: without explicit knowledge of the current
plaintext and/or the decomposition of $G$ to the matrices $A$ and $B$
it will require an exponentially long time to decipher a specific
ciphertext. Partial or complete knowledge of the ciphertext/plaintext
as well as partial knowledge of the matrix $B$ (while ${\cal O} (N)$
of the elements remain unknown) will not be helpful for decomposing
$G$ which will still require an exponentially long time to carry out.

We will consider here two attacks on specific plaintexts with partial
knowledge of the corrupting vector $\bz$ or of the matrix $B$. In the
first case, knowing $p_a M$ of the $p M$ corrupting bits may allow one
to subtract the approximated vector $\hz$ from the ciphertext and take
the product of $G^{-1}$ and the resulting ciphertext. This attack is
similar to the task facing an unauthorised user with a reduced
corruption rate of $(p\!-\! p_a)$. For any non-vanishing difference
between $p_a$ and $p$, deciphering the transmitted message remains a
difficult task.

A second attack is that whereby the matrix $B$ is known to some
degree; for instance, the location of a fraction of the unit elements,
say $1\!-\!\rho$ is known. From Eq.(\ref{eq:decoding}) one can
identify the absent information as having a higher effective
corruption level of $p \!+\! g(\rho) $, where $g(\cdot)$ is some
non-trivial function that depends on the actual scenario. To secure
the transmission one may work sufficiently close to the critical
corruption level $p_s$ such that the additional effective noise $\rho$
will bring the system beyond the critical corruption rate and into the
paramagnetic/spin-glass regime. However, the drawbacks of working {\em
very close} to $p_s$ are twofold: Firstly, average decryption times
using BP methods ($\tau$) will diverge proportionally to $1/(p_s \!
-\!p)$ as demonstrated in the inset of Fig.2. Secondly, finite-size
effects are expected to be larger close to $p_s$, which practically
means that the system may not converge to the attractive optimal
solutions in some cases.
  
We will end this presentation with a short discussion on the
advantages and drawbacks of the suggested method in comparison with
existing techniques. Firstly, we would like to point out the
differences between this method and the McEliece cryptosystem.  The
latter is based on Goppa codes and is limited to relative low
corruption levels. These may allow for decrypting the ciphertext using
(many) estimates of the corruption vector. Our suggestion allows for a
significant corruption level, thus increasing the security of the
cryptosystem. In addition, the suggested construction, $K\!\!=\!\!
L\!\!=\!\!  2$, is not discussed in the information theory literature
(e.g. in \cite{MacKay}) which typically prefers higher $K,L$ value
systems. Secondly, in comparison to RSA where decryption takes ${\cal
O} (N^3)$ operations, our method only requires ${\cal O} (N)$
operations, multiplied by the number of BP iterations (which is
typically smaller than 100 for most system sizes examined except very
close $p_s$).  Encryption costs are of ${\cal O} (N^2)$ (as in RSA)
while the inversion of the matrix $B$ is carried out only once and
requires $O(N^{3} )$ operations.

The two obvious drawbacks of our method are: 1) The transmission of
the public key, which is a dense matrix of dimensionality $M\!\times\!
N$. However, as public key transmission is carried out only once for
each user we do not expect it to be of great significance.  2) The
ciphertext to plaintext bit ratio is greater than one to allow for
corruption, in contrast to RSA where it equals 1.  Choosing the $N/M$
ratio is in the hands of the user and is directly related to the
security level required; we therefore do not expect it to be
problematic as the increased transmission time is compensated by a
very fast decryption.

We examine the typical performance of a new cryptosystem, based on
insight gained from our previous studies, by mapping it onto an Ising
spin system; this complements the information theory approach which
focuses on rigorous worst-case bounds.  We show that authorised
decryption is fast and simple while unauthorised decryption requires a
prohibitively long time. Important aspects that are yet to be
investigated include finite size effects and methods for alleviating
the drawbacks of the new method.

{\footnotesize
\vspace*{0.1in} {\bf \hspace*{-1.5em} Acknowledgement} Support by
JSPS-RFTF (YK), The Royal Society and EPSRC-GR/L52093 (DS) is
acknowledged. We would like to thank Manfred Opper and
Hidetoshi Nishimori for reading the manuscript.}

\newpage

\begin{figure}[h]
\vspace*{-1.25cm}
\begin{center}
\begin{picture}(400,400)
\put(-40,200){\epsfxsize=100mm  \epsfbox{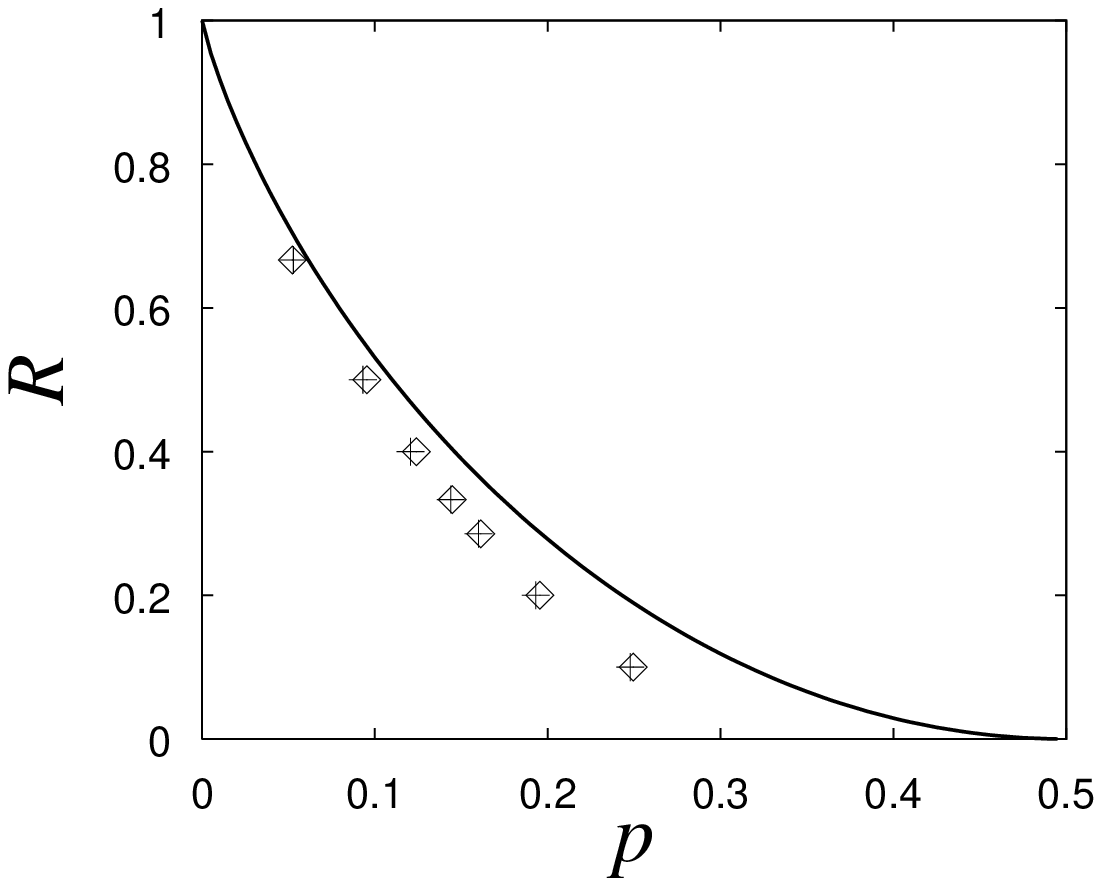}}
\put(115,270){\epsfxsize=40mm  \epsfbox{PT_K2L2.eps}}
\end{picture}
\end{center}
\vspace*{-7.3cm}
\caption{Critical transmission rate as a function of the corruption
rate $p$ for an unbiased ciphertext. Numerical solutions (of the
analytically obtained equations - $\Diamond$) and BP iterative
solutions (of system size $N\!=\!  10^4$, $+$), were averaged over 10
different initial conditions of almost zero magnetisation with error
bars much smaller than the symbol size.  Inset: The ferromagnetic
({\sf F}) (optimal/sub-optimal) and paramagnetic ({\sf P}) solutions
as functions of $p$; thick and thin lines denote stable solutions of
lower and higher free energies respectively, dashed lines correspond
to unstable solutions.}
\end{figure}


\begin{figure}[h]
\vspace*{-2.35cm}
\begin{center}
\begin{picture}(400,400)
\put(-20,200){\epsfxsize=95mm  \epsfbox{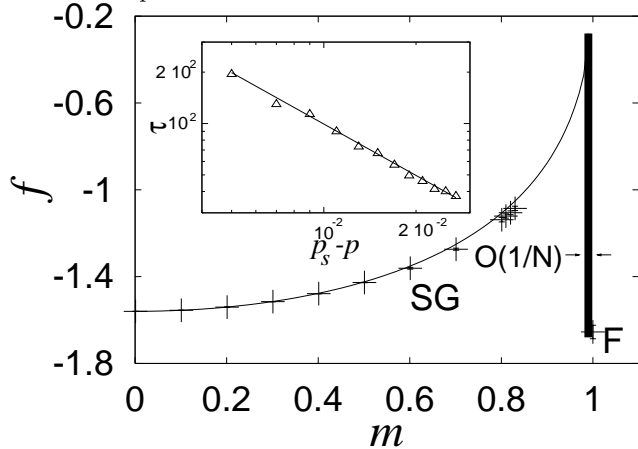}}
\end{picture}
\end{center}
\vspace*{-7.6cm}
\caption{The free energy landscape as a function of $m$ for the
transmission rate $N/M=1/2$ and flip rate $p=0.08$; theoretical values
are represented by the solid line, numerical data, obtained by BP
($N=200$) and averaged over 10 different initial conditions, are
represented by symbols ($+$).  The landscape is deep and narrow (of
width ${\cal O} (1/N)$) at $m=1$ and rather flat elsewhere.  Inset -
scattered plot of mean decryption times - $\tau$. The optimal fitting
of straight lines through the data provides another indication for the
divergence of decryption time for corruption rate close to
$p_s=0.953\pm5$ (in this example).}
\end{figure}

\end{document}